# *Ab initio* probing of the electronic band structure and Fermi surface of fluorine-doped $WO_3$ as a novel low-$T_C$ superconductor

I.R. Shein,\*  A.L. Ivanovskii

*Institute of Solid State Chemistry, Ural Branch of the Russian Academy of Sciences, 620990, Ekaterinburg, Russia * shein@ihim.uran.ru*

First-principles calculations were performed to investigate the electronic structure and the Fermi surface of the newly discovered low-temperature superconductor: fluorine-doped $WO_3$. We find that F doping provides the transition of the insulating tungsten trioxide into a metallic-like phase $WO_{3-x}F_x$, where the near-Fermi states are formed mainly from W $5d$ with admixture of O $2p$ orbitals. The cooperative effect of fluorine additives in $WO_3$ consists in change of electronic concentration as well as the lattice constant. At probing their influence on the near-Fermi states separately, the dominant role of the electronic factor for the transition of tungsten oxyfluoride into superconducting state was established. The volume of the Fermi surface gradually increases with the increase of the doping. In the sequence $WO_3 \rightarrow WO_{2.5}F_{0.5}$ the effective atomic charges of W and O ions decrease, but much less, than it is predicted within the idealized ionic model - owing to presence of the covalent interactions W-O and W-F.

Electron or hole doping of solids via atomic substitutions provides wide and diverse possibilities for tuning their physical properties and opens up fascinating ways to control their functionalities. Today, for promising groups of materials, series of dopants, which can exert the expected influence on their carrier concentration, are an issue of extensive discussions.

So, for oxides, nitrogen and carbon, having one or two electron less than oxygen, may be viewed as "universal" hole dopants. The introduction of these dopants into insulating oxides can induce a rich set of unexpected physical phenomena, for example, so-called $d^0$-magnetism [1], which offers new opportunities for search of high-temperature spintronic materials, see [2-4].

In turn, the injection of electrons also can cause quite non-trivial changes in the properties of oxides. Here, fluorine with one additional electron compared to oxygen may be considered as a "universal" electron dopant. The most widely known example, which has triggered enormous scientific activity in the field of so-called Fe-based high-temperature superconductors (HTSCs) in the last three years (reviews [5-8]), is metallic-like LaFeAsO, for which superconductivity (with $T_C \sim 26K$) emerges [9] from the anti-ferromagnetic ordered ground state upon fluorine doping.

Very recently, it was reported [10] that superconductivity can also be induced by fluorine doping in simple insulating oxides such as $WO_3$. In experiments [10], the cubic tungsten oxyfluorides $WO_{3-x}F_x$ were prepared by fluorination of $WO_3$ using polytetrafluoroethylene (Teflon); for the synthesized samples with fluorine contents of $0.41 < x < 0.45$, low-temperature superconductivity (LTSC, $T_C \sim 0.4K$) was discovered.

Generally, two main effects of substitution F → O in $WO_3$ can be responsible for tuning of electronic properties from insulating into superconducting state, namely, carrier doping and a steric effect caused by the difference in ionic radii of fluorine $R(F^-) = 1.29$ E and oxygen ions $R(O^{2-}) = 1.35$ E.

In this Letter, by means of the first principles band structure calculations, we explore the electronic structure and the Fermi surface (FS) topology of the novel LTSC: fluorine-doped $WO_3$ and discuss how they depend on the above mentioned electronic and structural factors.

The considered $WO_{3-x}F_x$ adopts the cubic $ReO_3$-like structure with space group *Pm-3m*, where the Wyckoff positions of atoms are tungsten at *1a*: (0, 0, 0) and oxygen (fluorine) at *3d*: (S, 0, 0). In our calculations, the periodic 32-atomic supercell 2×2×2 ($W_8O_{24}$) in cubic $WO_3$ was used; then, with partial replacing O → F, the supercell $W_8O_{20}F_4$ was constructed to simulate the tungsten oxyfluoride with the formal stoichiometry $WO_{2.5}F_{0.5}$. Further this cell was converted in orthorhombic space group *Fmm*2 - $W_4O_{10}F_2$ with lattice parameters $a' = 2a, b' = c' = 2\sqrt{2}a$.

The calculations are performed by means full-potential linearized augmented plane wave method (FPLAPW) method implemented in the WIEN2k suite of programs [11]. At the first step, the generalized gradient approximation (GGA) for the exchange-correlation potential in PBE form [12] was used. Secondly, the calculations are performed with a fully relativistic treatment of the core states and a scalar-relativistic treatment including spin-orbit coupling (SOC) for the valence states, as implemented in WIEN2k.

The core states were treated within the spherical part of the potential only and were assumed to have a spherically symmetric charge density confined within MT spheres. The valence part was treated with the potential expanded into spherical harmonics to $l = 4$. The plane-wave expansion with $R_{MT}$ Ч $K_{MAX}$ was equal to 7.0, and *k* sampling with 6Ч6Ч6 *k*-points mesh in the full Brillouin zone was used. The MT sphere radii were chosen to be 1.85 a.u. for W, and 1.6 a.u. for O, and for F. In our calculations of fluorine-doped $WO_3$, the lattice parameter $a = 3.81219$ Å [10] was used.

In Fig. 1 we present calculated total, atomic, and orbital decomposed partial densities of states (DOSs) for $WO_{2.5}F_{0.5}$. We see that after the partial substitution of fluorine for oxygen, owing to the growth of electronic concentration (*EC*), the Fermi level ($E_F$) is shifted upward (compared with insulating $WO_3$ [13-17]) into the bottom of conduction band, leading to the metallic-like behavior of cubic $WO_{2.5}F_{0.5}$. The electronic states on the Fermi level and around it (from -1 eV to +2 eV) are mostly formed by W 5*d* orbitals, whereas the O 2*p* orbitals are in the range from -2 eV up to -11.5 eV. Valence states of fluorine are placed at the bottom of mixed W 5*d* - O 2*p* band.

Comparing the DOSs shapes as obtained at *non*-SOC and SOC levels, we find that the most significant differences are observed in the states of valence band ranging from -2 eV up to -11.5 eV, Fig. 1. On the contrary the changes of the near-Fermi states appear quite insignificant: so, the values of total DOS at the Fermi level $N^{tot}(E_F)$ as calculated with and without SOC differ at about 6%, see Table. Using the experimentally derived electronic specific coefficient $\gamma = 1.59$ mJ·$K^{-2}$·$mol^{-1}$ (for the sample with composition $WO_{2.56}F_{0.44}$ [10]) within the free electron model we have estimated the value of $N^{tot}(E_F)^{exp} \sim 0.57$ 1/eV·form.unit, which became in general comparable with calculated data, see Table. The discrepancy between our estimated values and experimental data should be attributed mainly to known shortcomings of the DFT calculations as well as to differences in compositions of $WO_{3-x}F_x$ - in our model calculations and for experimentally examined sample.

In Fig. 2 we present the Fermi surface of $WO_{2.5}F_{0.5}$. The common FS adopts a 3D-like topology and comprises three enclosed sheets. Among them the external sheet is formed by interconnected deformed cylinders extended along the $k_x$, $k_y$, and $k_z$ directions. The second sheet consists of three basic disconnected cuboid-like pockets, which are centered at the $\Gamma$ point and at the top and bottom sides of the Brillouin zone (BZ); besides, two small closed pockets are in the corners of BZ. The third sheet comprises three interconnected ellipsoidal pockets centered at the $\Gamma$ and $X$ points with very small pockets at the edges of BZ, Fig. 2.

Next, since the main contributions to the near-Fermi region come from tungsten and oxygen states, the effect of fluorine additives in $WO_{3-x}F_x$ consists mainly in the change of the *EC* and the lattice constant. In order to evaluate the influence of these factors on the near-Fermi states separately, two independent estimations were made.

Firstly, using the band picture as obtained for the given experimental lattice parameter, we will vary the *EC*, simulating (in the framework of the rigid-band model) the doping level in $WO_{3-x}F_x$ in the range $0.2 < x < 0.7$.

Secondly, using the fixed composition $WO_{2.5}F_{0.5}$ (*i.e.* fixed *EC*), we will vary the lattice constant. Here, the empirical approximation [10] $a = 3.792 + 0.045x$ for the same range of x ($0.2 < x < 0.7$) was used.

For both cases we calculated the changes in $N^{tot}(E_F)$, Fig. 3. We find that the increase of the *EC* (*i.e.* growth of doping level at fixed lattice constant) is accompanied by an essential growth of the density of states at the Fermi level; on the contrary, in case of lattice expansion (at fixed *EC*) this effect is quite small. Numerically, the increase in $N^{tot}(E_F)$ in the range $0.2 < x < 0.7$, $\delta N^{tot}(E_F) = \{N^{tot}(E_F)^{x=0.7} - N^{tot}(E_F)^{x=0.2}\}/N^{tot}(E_F)^{x=0.2}$, is $\delta N^{tot}(E_F) \sim 3\%$ when only the lattice parameter is changed, whereas with a change in the *EC*, $\delta N^{tot}(E_F) \sim 76\%$. Hence, we conclude that with a growth of the F/O ratio, the main factor responsible for the experimentally observed metallization of fluorine-doped $WO_3$ and subsequent transition of $WO_{3-x}F_x$ into superconducting state (in the assumption of the BCS pairing model, where $T_C \sim N^{tot}(E_F)$) is electron doping.

Let us consider the evolution of the Fermi surface topology across the doping level. The Fermi surface is associated with a change in the band occupancy, and in Fig. 2 we represent the FSs for the lowest (x=0.2) and the highest (x=0.7) doping levels considered. We see that with an increase in the band occupancy, the FSs retain their shape in the doping interval $0.2 < x < 0.5$, and some visible reconstruction of the FSs occurs for $x > 0.5$. In general, the volume of the Fermi surface gradually increases with the *EC* (doping level). Note that a similar trend was found for the related system: sodium tungsten bronze $Na_xWO_3$, where Na ions intercalated into $WO_3$ lattice behave as electron dopants. Here, by means of the high-resolution angle-resolved photoemission spectroscopy (HR-ARPES), the Fermi surface was found [18,19] to increase with Na concentration (*i.e.* with an increase in the electron doping level).

Finally, the influence of fluorine additives on the atomic charges in $WO_3$ is of interest. So, the increase in the lattice constant of tungsten oxyfluoride with a growth of F content (in spite of the fact that $R(F^-) < R(O^{2-})$) is explained [10] by a transition of a part of ions $W^{6+}$ into $W^{5+}$ with a larger radius during fluorination. We carried out a Bader [19] analysis, and the calculated effective atomic charges Q for ideal $WO_3$ are $Q(W) = +5.01$ and $Q(O) = -1.67$. The corresponding averaged charges for $WO_{2.5}F_{0.5}$ are $Q(W) = +4.29$, $Q(O) = -1.55$, and $Q(F) = -0.83$. Thus in the sequence $WO_3 \rightarrow WO_{2.5}F_{0.5}$ the effective atomic charges decrease, but in general the inter-atomic charge transfer is much smaller than it is predicted in the aforementioned idealized ionic model. This can be explained by the peculiarities of inter-atomic bonding in $WO_3$ [13-17] and $WO_{2.5}F_{0.5}$, which is of a mixed ionic-covalent type, when, besides ionic bonding, moderate covalent W-O (W-F) interactions arise owing to hybridization of W $5d$ - O $2p$ (W $5d$ - F $2p$) states, see Fig. 1.

In summary, the electronic properties and the Fermi surface of the newly discovered LTSC: fluorine-doped $WO_3$ were examined within first principles band structure calculations. We find that the near-Fermi states in $WO_{3-x}F_x$ are formed mainly from W $5d$ and O $2p$ orbitals, which are responsible for the transport properties of this material; the influence of spin-orbit coupling on these states is quite insignificant. Fluorine additives in $WO_3$ lead to the changes in the electronic concentration and the lattice constant. The results obtained emphasize the dominant role of the electronic factor providing the transition of tungsten oxyfluoride into superconducting state. Our calculations also revealed that some changes in the shape of the FS will take place with doping, while the volume of the Fermi surface in general gradually increases with an increase in the doping level (F/O ratio). Finally, fluorine doping provides some reduction of effective atomic charges of $W^{n+}$ and $O^{m-}$ ions, but these changes are much less than it is expected within ideal ionic model – owing to covalent contributions into overall bonding picture.

---

**Total and partial densities of states at the Fermi level ($N^{tot}(E_F)$, in 1/eV·form. unit) for $WO_{2.5}F_{0.5}$ as calculated within non-SOC and SOC levels.**

| states | $N^{tot}(E_F)$ | $N^{W5d}(E_F)$ | $N^{O2p}(E_F)$ | $N^{F2p}(E_F)$ |
|---|---|---|---|---|
| *non*-SOC | 0.775 | 0.363 | 0.109 | 0.011 |
| SOC | 0.823 | 0.389 | 0.107 | 0.013 |

**FIGURES**

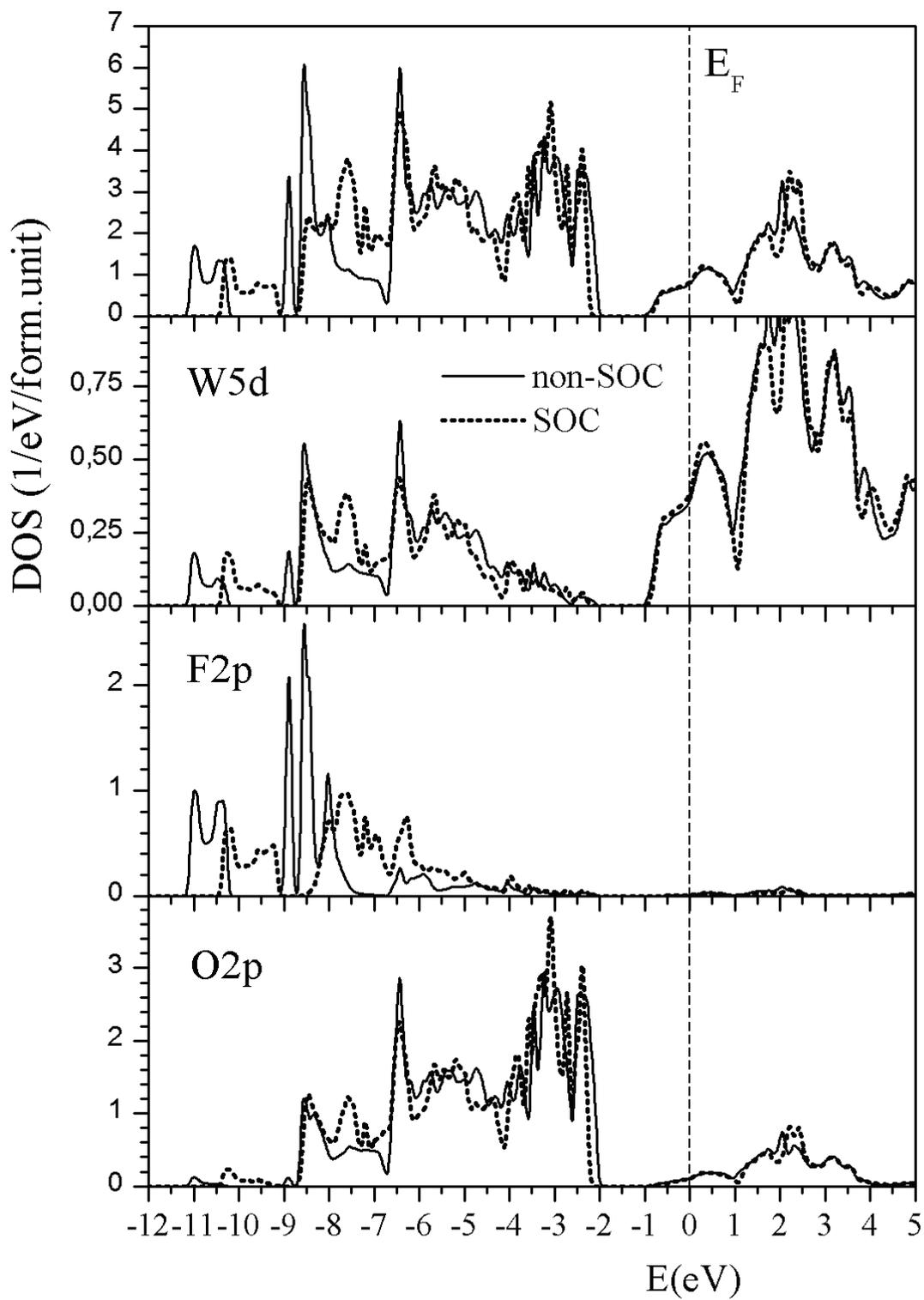

Fig. 1. Total (*upper panel*) and partial densities of states for $WO_{2.5}F_{0.5}$ as obtained at *non*-SOC and SOC levels

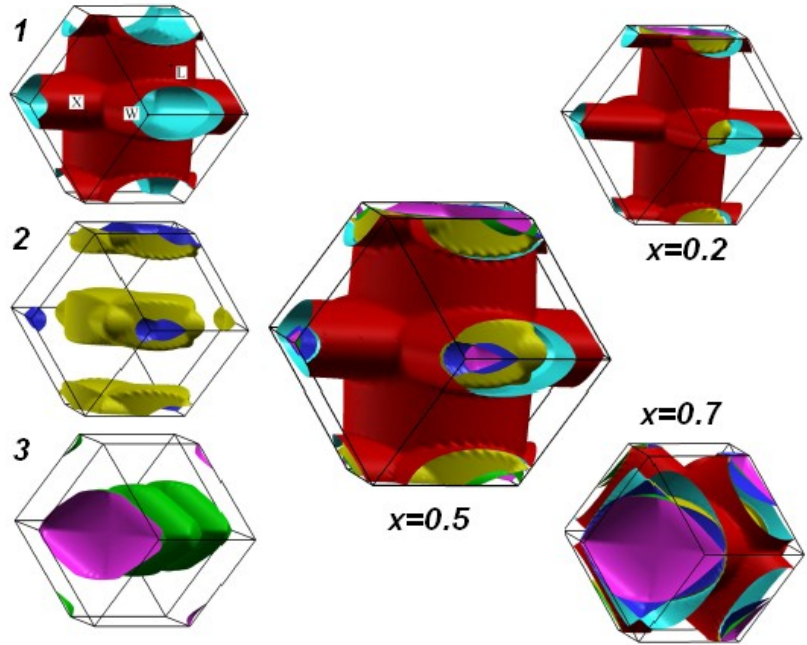

Fig. 2. Separate sheets of the Fermi surface for $WO_{2.5}F_{0.5}$: *1-3* (*at the left*); common FS for $WO_{2.5}F_{0.5}$ (*in the centre*); and the FS evolution for $WO_{3-x}F_x$ for different doping levels: $x = 0.2$ and $x = 0.7$ (*at the right*).

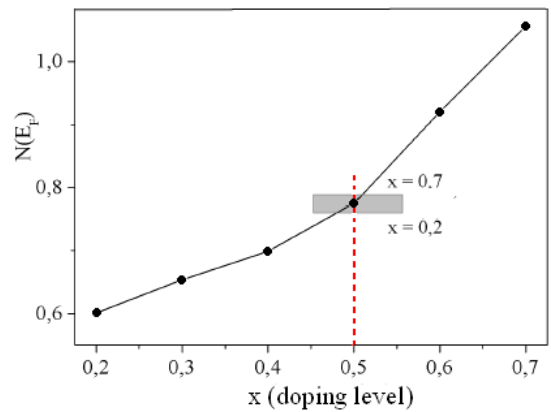

Fig. 3. Total densities of states $N^{tot}(E_F)$ (in 1/eV·form.unit) of $WO_{3-x}F_x$ as a function of x (doping level). For the composition $WO_{2.5}F_{0.5}$, the interval of $N^{tot}(E_F)$ changes is given as a function of the lattice parameter *a* (approximated in the range $0.2 < x < 0.7$, *see the text*).